\documentstyle[12pt,a4]{article}
\def\be{\begin{equation}}
\def\ee{\end{equation}}
\def\bea{\begin{eqnarray}}
\def\eea{\end{eqnarray}}

\def\nn{\nonumber}

\def\prd{Phys. Rev. \underline}
\def\pl{Phys. Lett. \underline}

\def\beq{\begin{equation}}
\def\eeq{\end{equation}}
\def\noi{\noindent}
\def\bea{\begin{eqnarray}}
\def\eea{\end{eqnarray}}
\def\bei{\begin{itemize}}
\def\eei{\end{itemize}}

\def\fourthrm{\font\fourthrm=cmr12 at 14pt}

\def\bal{{\bar \alpha}}
\def\bet{{\bar \beta}}
\def\eqtipm{ {\,\,\lower.7ex\hbox{$=\atop T\rightarrow\pm\infty$}\,\,}}

\begin{document}

\begin{center}
\title{\bf Factorization versus Duality in Nonleptonic decays,\\
a Quark Model Approach.}

\author{ A. Le Yaouanc, L. Oliver, O. P\`ene and J.-C. Raynal} \par
\maketitle
{Laboratoire de Physique Th\'eorique et Hautes Energies\footnote{Laboratoire
associ\'e au
Centre National de la Recherche Scientifique - URA D00063}}\\
{Universit\'e de Paris XI, B\^atiment 211, 91405 Orsay Cedex, France}\\
\end{center}
\vskip 5 mm
\noindent {\bf Abstract}\\
\begin{abstract}
We study in a quark model the contradiction between factorization and duality
found in nonleptonic decays at next to leading order in $1/N_c$, concentrating
on quark exchange mechanism. The contradiction originates in the fact that the
standard factorization assumption approximates the asymptotic final states by a
non-orthogonal set of states, thus leading to an overcounting of the decay
probability. We consider a system with two heavy quarks treated as classical
color sources with constant velocity, and two mass-degenerate antiquarks.
Exploiting permutation symmetry in an adiabatic approximation, we find that
final state interaction restores duality. Three $O(1/N_c)$ effects are
exhibited: i) a proper treatment of orthogonality yields a global correction
$1/N_c\to 1/2N_c$  within a generalized factorization \`a la BSW, (such a
factor was present in an Ansatz by Shifman),  ii) the distorsion of the meson
wave functions at the time of the weak decay, iii) relative phases generated by
the later evolution. The latter effect becomes dominant for light antiquarks or
for a small velocity of the final mesons, and may thoroughly modify the
factorization picture. For exclusive
decay it may interchange the role of class I and class II final channels, and
for semi-inclusive decay it may lead to an equal sharing of the probability
between the two sets of final states. In the heavy antiquark and large velocity
limit, the replacement $1/N_c\to 1/2N_c$ is the dominant correction.

\end{abstract}
\begin{flushright}
LPTHE-Orsay 95/26,\\
hep-ph/9504262
\end{flushright}
\newpage

\section{Introduction.}

While some progress has been performed during these last years in our
understanding of semileptonic and leptonic decay mechanisms, our understanding
of nonleptonic decays is still semi-quantitative. Not to speak of the $\Delta
I=1/2$ mystery, the non-leptonic decays of $D$ and $B$ mesons are most often
studied with the help of the standard factorization assumption \cite{svz}, the
theoretical basis for which exists only in the $N_c\to \infty$ limit, or of the
generalized factorization assumption \`a la BSW \cite{bsw1} which is a
phenomenological Ansatz.

The nonleptonic decay channels of heavy meson are an important issue and will
grow even more so, since they provide the cannels in which the CP asymmetries
will be looked for in $B$-factories.

 A critical study of factorization assumption is an urgent task, and has indeed
been started \cite{milana}, at a time when  increasingly accurate experimental
results learn us that our present understanding
of $B\to \psi K(K^\ast)$, based on factorization assumption among other
hypotheses, severely fails \cite{bpsik}.

It is usually claimed that the corrections to factorization are due to final
state interaction (FSI). In a sense this statement is true, but the proper
meaning of what is understood by that needs clarification.
 One of our aims in this paper is to proceed to this clarification in a simple
model in which the dynamics is rather transparent.

Furthermore little is known about the validity of factorization except that it
is  violated in low energy $K\to \pi\pi$ and  $D\to K\pi, KK, \pi\pi$
\cite{pham} \cite{kamal} channels where
strong FSI phases are experimentally known to be present, both from the direct
analysis of weak nonleptonic decays and from scattering experiments. The study
grows more difficult at larger energies when more channels are coupled. This
happens when multiparticle channels come in, particularly multi-pion channels,
and also when several two-body channels communicate
via strong interaction through quark exchange or quark-pair
annihilation-creation.

We will concentrate on two-body channels communicating via quark exchange. In a
Tamm-Dancoff type expansion, quark exchange is dominant since it needs no quark
pair creation neither annihilation. Furthermore, it has been stressed by J.
Donoghue \cite{donoghue} that such a mechanism might explain the $D^0\to\phi
\overline K^0$ decay amplitude.

On the other hand, M. Shifman \cite{shifman} (see also \cite{alek18}) has made
the interesting remark that beyond leading $1/N_c$ order, the factorization
assumption simply violates duality. As we shall show at length in this paper,
this effect is fully related to final state interaction via quark exchange.

Indeed this comes from the fact that in color space a $q q \bar q \bar q$ color
singlet
can be decomposed in two ways into two $q \bar q$ color singlets, but the two
resulting states are not orthogonal to each other \cite{alek18}. At the time
the weak decay of the initial meson creates a $q q \bar q \bar q$ color
singlet, the four quarks interact strongly with one another. It is impossible
to tell that one pair $q \bar q$ is in one meson and the other pair in another
meson. However the system evolves, and eventually splits into two  $q \bar q$
color singlets that are spatially distant and hence orthogonal. During all the
period of the interaction, one given quark does not know with which antiquark
it is paired. This is exactly the situation which is expressed usually by the
expression ``quark exchange''. What is depicted by this expression is not a
simple and instantaneous exchange of a quark from one meson to another, but a
period of overlap of the two mesons resulting finally in a non-vanishing
amplitude for an exchange of a quark.

The factorization assumption totally overlooks this complex mechanism, as it
simply computes the overlap of the $q q \bar q \bar q$ system resulting from
weak decay with the final mesons. {\it Two non-orthogonal states are taken as
an approximation of two distinct final states which are obviously orthogonal}.
Once one realizes this, it is not a surprise that one encounters some problems
with probability conservation.

Our aim is mainly to understand better this interaction mechanism in a simple
model, to check that when the dynamics is correctly treated there is no
contradiction with duality, and to identify the different effects contributing
at next to leading order in $1/N_c$.

We will work in the kinematical situation considered by Shifman \cite{shifman},
in a Hamiltonian approach, namely a quark model one. In section 2. we will
rephrase duality in the language of closure theorem and reexpress the
contradiction between factorization and duality. In section 3 we will present a
quark model, with an adiabatic approximation and mass-degenerate antiquarks,
and we exploit the resulting permutation symmetry to compute the S matrix and
the weak decay amplitude. In section 4 we will conclude.

\section{Duality versus factorization, rephrasing the problem. }

Let us recall the ideal process which is studied in \cite{shifman}.
There are three heavy quarks $A, B, C$ and two light antiquarks
$\bal, \bet$. We will assume the latter to be heavy enough to
 justify the use of the quark potential model which will be
our tool all along.

The process under study is the weak decay

\[ P_{A\bal} \to M_{B\bal}+M_{C\bet},\,\,M_{B\bet}+M_{C\bal}\]
where $P_{A\bal}$ is pseudoscalar meson composed by $A$ and $\bal$
and
$M_{B\bal}$ represents any meson composed by $B$ and $\bal$.

The following relations are assumed \cite{shifman}:

\be M_B=M_C\equiv M,\quad M_A=2M+\Delta,\quad
\Lambda_{QCD} \ll \Delta\ll M \label{m1}\ee

To which we add
\be \Lambda_{QCD}\ll m_{\alpha}, m_{\beta}  \label{m2}\ee
to justify the use of quark model.

We cannot find a physical example of such a situation.
If the $s$ quark was heavy, the ideal situation assumed
in \cite{shifman} would be realized by the couple of decay channels:
$B_d\to D^0\overline K^0, D^+ K^-$.

\subsection{Relation between duality and closure theorem.}

In \cite{shifman} M. Shifman exhibits a contradiction
between duality and the standard factorization hypothesis also encountered
in \cite{alek18} while studying $\Delta \Gamma$ for the $B_s-\overline B_s$
system. We will study
this issue as a contradiction between the closure theorem
and the standard factorization hypothesis. Indeed, duality is related
to the closure theorem in quantum mechanics as we will now
recall.

Let us call generically $|n>$ all hadronic states built up with
quarks $B, C, \bal, \bet$. Calling $H_W$ the weak Hamiltonian,
the state $H_W| P_{A\bal}>$  is composed of the four quarks
$B, C, \bal, \bet$. The decay width of the $P_{A\bal}$ meson is
given by

\bea \Gamma	(P_{A\bal})=&\sum_n
<P_{A\bal}|H_W|n><n|H_W|P_{A\bal}>\delta(E_n-E_{A\bal})\nn\\
=& \sum_n <P_{A\bal}|H_W|n>\delta(H-E_{A\bal})<n|H_W|P_{A\bal}>\label{dualite}
\eea
where $E_{A\bal}$ is the initial energy, $E_n$ is the energy of the
state $|n>$, $H$ is the strong Hamiltonian ($H|n>=E_n|n>$), and where
the sum is to be understood as a sum over discrete states and an integral
over continuum states. The set of states $|n>$ is a complete set. The sum could
be
expanded on any basis, and in particular on the basis of the free quarks
$B, C, \bal, \bet$. This is where the closure of the Hilbert space comes in.

Now, $H=H_c+V$ where $H_c$ stands for the kinetic energy and $V$ for the
potential.
Whenever the contribution from $V$ can be neglected in front of $H_c$, one
recovers lowest order duality, i.e. the simple parton model with no
perturbative corrections neither non-perturbative ones from higher dimension
operators. We do not want to go into the question of when this approximation is
valid, and how to get better approximations.  We simply want to rephrase the
contradiction between factorization and duality as a contradiction between
factorization and closure, then go into a simple model to show how the dynamics
solves this.

\subsection{Contradiction between factorization and closure.}

Let us consider for example the weak Hamiltonian:
\be H_W = 2 {\sqrt 2} G  \left[C_1 (\overline B \gamma_\mu L A)(\overline C
\gamma^\mu L\beta)
+C_2  (\overline B \gamma_\mu L\beta) (\overline C \gamma^\mu L A)
\right]\label{wh}\ee
where $L=(1-\gamma_5)/2$, $C_1$ and $C_2$ are coefficients that we do not need
to specify in this paper, although they are reminiscent of the familiar
coefficients in the effective weak interaction Hamiltonian.

At the time $t=0$ the weak Hamiltonian acts on the initial
meson and produces a state that contains the quarks $C, \bet, \bal, B$. Let us
call this state $|f>$. We find it convenient for later use to decompose $|f>$
into its color part and the remainder:

\begin{eqnarray}\lefteqn {|f>\equiv H_W |P_{A\bal}> =
\sum_{\begin{array}{l}\tiny{s_C, s_\bet, c_C, c_\bet,}\\ \tiny
{s_B,s_\bal,c_B,c_\bal}\end{array}}
\int d\vec p_C d\vec p_\bet d\vec p_B d\vec p_\bal \Psi(\vec p_C, s_C, \vec
p_\bet, s_\bet,\vec p_B, s_B, \vec p_\bal, s_\bal)}\nn\\  &
\frac 1 {N_c} [C_1 \delta_{c_C,c_\bet}\delta_{c_B,c_\bal}
+  C_2 \delta_{c_C,c_\bal}\delta_{c_B,c_\bet}]\,
|C,\vec p_C, s_C, c_C ; \bet, \vec p_\bet, s_\bet,
c_\bet;B,\vec p_B, s_B, c_B ; \bal, \vec p_\bal, s_\bal,
c_\bal > \label{f}\eea
where $\vec p_C, s_C, c_C$ ($\vec p_\bet, s_\bet, c_\bet$) labels the momentum,
spin and color of the (anti-)quark $C$ ($\bet$), $N_c$ is the number of colors.
The function $\Psi$ may be computed from the wave function of $P_{A\bal}$
and the operator $H_W$ in (\ref{wh}). However we will skip this computation
since the precise expression for $\Psi$ is not relevant for our argument.

It is obvious that:
\be
<f|f>= \left(C_1^2+C^2_2 + 2 \frac {C_1 C_2}{N_c}
\right) K \label{ff}\ee
with
\be K=\sum_{\begin{array}{l}\tiny{s_C, s_\bet, c_C, c_\bet,}\\ \tiny
{s_B,s_\bal,c_B,c_\bal}\end{array}}
\int d\vec p_C d\vec p_\bet d\vec p_B d\vec p_\bal\left\vert \Psi(\vec p_C,
s_C, \vec p_\bet, s_\bet,\vec p_B, s_B, \vec p_\bal, s_\bal)\right\vert^2
\label{K}\ee

Let us also decompose the meson wave function into a color part and the
remainder:
\be |M^{(n)}_{C\bet}>= \sum_{s_C,s_\bet,c_C,c_\bet}\int d\vec p_C d\vec p_\bet
\psi^{(n)}_{C\bet}(\vec p_C, s_C, \vec p_\bet, s_\bet)\frac 1{N_c^{\frac 1 2}}
\delta_{c_C,c_\bet}|C,\vec p_C, s_C, c_C ; \bet, \vec p_\bet, s_\bet,
c_\bet>\label{meson}\ee
and analogously for all quark-antiquark pairs. The $\psi^{(n)}_{C\bet}$'s form
a complete orthonormal basis of the spin-momentum $C\bet$ Hilbert space.
Let us now define the spin-space overlaps:

\bea K_{C\bet;B\bal}^{(n,m)}=& \sum_{s_C, s_\bet,s_, s_\bal}\int d\vec p_C
d\vec p_\bet
d\vec p_B d\vec p_\bal\,\,
\Psi^\dagger(\vec p_C, s_C, \vec p_\bet, s_\bet,\vec p_B, s_B, \vec p_\bal,
s_\bal) \nn \\
&\psi^{(n)}_{C\bet}(\vec p_C, s_C, \vec p_\bet, s_\bet)
\psi^{(m)}_{B\bal}(\vec p_B, s_B, \vec p_\bal, s_\bal)\label{overlap}\eea
and analogously for the alternative grouping of quark-antiquark pairs:
$C\bal;B\bet$. Closure in the  $C\bet$ and $B\bal$ spin-momentum subspaces
implies that

\be \sum_{n,m}  K_{C\bet;B\bal}^{(n,m)\ast}K_{C\bet;B\bal}^{(n,m)}
=  K,\label{cl1}\ee
and analogously
\be \sum_{n,m}  K_{B\bet;C\bal}^{(n,m)\ast}K_{B\bet;C\bal}^{(n,m)}
=  K.\label{cl2}\ee

{}From eqs. (\ref{f}), (\ref{meson}) and (\ref{overlap}) it results that

\bea <f|M^{(n)}_{C\bet}; M^{(m)}_{B\bal}>=& a_1 K_{B\bet;C\bal}^{(n,m)}
\nn\\
<f|M^{(n)}_{B\bet}; M^{(m)}_{C\bal}>=& a_2 K_{B\bet;C\bal}^{(n,m)}
.\label{over2}\eea
with
\be a_1=C_1+\frac {C_2} {N_c},\qquad a_2=C_2+\frac {C_1} {N_c}\label{a1a2}\ee

Up to now all equations were exact. Now we shall formulate in our formalism the
factorization approximation by assuming
 that the decay amplitudes are well approximated by the overlaps:

\bea T\left(P_{A\bal}\to M^{(n)}_{C\bet} M^{(m)}_{B\bal}\right) \simeq
<f|M^{(n)}_{C\bet} M^{(m)}_{B\bal}>^\ast\nn\\
T\left(P_{A\bal}\to M^{(n)}_{B\bet}; M^{(m)}_{C\bal}\right) \simeq
<f|M^{(n)}_{B\bet} M^{(m)}_{C\bal}>^\ast\label{factorization}\eea

More precisely, the standard factorization assumption \cite{svz} uses eqs.
(\ref{over2}) (\ref{a1a2}) and (\ref{factorization}) with $C_1$, $C_2$ computed
from the electroweak theory complemented with QCD radiative corrections. Bauer,
Stech and Wirbel \cite{bsw1} have proposed a phenomanological factorization
assumption that keeps eqs. (\ref{over2}) and (\ref{factorization}) but
with $a_1$ and $a_2$ fitted to all known $D$, respectively $B$, decays.

Within standard factorization, summing over all two meson final states and
using eqs.
(\ref{f}), (\ref{K}), (\ref{cl1}), (\ref{cl2}), (\ref{over2}) (\ref{a1a2}) and
(\ref{factorization}) we get:

\be \sum_{n,m}\vert T\left(P_{A\bal}\to M^{(n)}_{C\bet}
M^{(m)}_{B\bal}\right)\vert^2 + \vert T\left(P_{A\bal}\to M^{(n)}_{B\bet}
M^{(m)}_{C\bal}\right)\vert^2= \left((C_1^2 +C^2_2)(1+\frac 1 {N_c^2}) + 4\frac
{C_1C_2} {N_c} \right) K\label{paradoxe}\ee

To leading order in $1/N_c$ eq. (\ref{paradoxe}) gives the same result as
eq. (\ref{ff}). In our present framework this reflects the well known fact that
to leading order in $1/N_c$ factorization and duality are compatible. However,
the $O(1/N_c)$ corrections show a discrepancy, the contradiction stressed in
\cite{shifman}. As suggested in  \cite{shifman}, this discrepancy could be
cured, except for the $1/N_c^2$ terms  by using a phenomenological
factorization with
\be a_1=C_1 + \frac {C_2} {2 N_c},\qquad a_2=C_2 +\frac {C_1} {2
N_c}\label{a1a2s}\ee

However, in our framework it is easy to trace back the origin of the
discrepancy. The fact is that the set of states $|M^{(n)}_{C\bet}
M^{(m)}_{B\bal}> \oplus |M^{(n)}_{B\bet} M^{(m)}_{C\bal}>$ is not an
orthonormal basis of the Hilbert space. The states are normalised, but they are
not orthogonal:

\be <M^{(n)}_{C\bet} M^{(m)}_{B\bal}|M^{(n)}_{B\bet} M^{(m)}_{C\bal}>
=O(\frac 1 {N_c})\ne 0\label{ne0}\ee
in general. This overlap is $O(1/N_c)$ as can be easily derived from the color
part in wave function  (\ref{meson}):
\be \left (\frac 1 {N_c^{\frac 1 2}} \right)^4
\sum_{C_C,C_B,C_\bal,C_\bet} \delta_{C_C,C_\bet}\delta_{C_B,C_\bal}
\delta_{C_B,C_\bet}\delta_{C_C,C_\bal}= \frac 1 {N_c}\label{colap}\ee
It is easy to check that (\ref{ne0}) is at the origin of the discrepancy
between (\ref{paradoxe}) and (\ref{ff}).

Neither are these states eigenstates of the strong Hamiltonian. Indeed, these
states are built up from two asymptotic mesons combined via a plane wave for
the relative momentum between the two mesons. When the two mesons lie far
apart, the simple product of their wave function is an eigenstate of the strong
Hamiltonian. However, in the states we consider
there is a non negligible contribution with the two mesons overlapping in space
where they strongly interact leading to an important distortion from the simple
product of asymptotic meson wave functions.

\section{An Adiabatic Quark Model with Degenerate Antiquarks.}

\subsection{The S matrix}

Let us first consider an oversimplified model. We will assume all quarks to be
spinless and $\bal$ and $\bet$ to be degenerate in mass: $m_\bal=m_\bet\equiv
m$. Next, $B$ and $C$ being very heavy, we will treat their motion as
classical. They are supposed to move head-on with velocity $\vec v$. As a
function of time $t$ the
spatial coordinates of $B$ and $C$ are
\be
\vec r_B=-\vec r_C= \vec v t.\label{class} \ee

{}From now on the mesons will be assumed to have their center-of-mass localised
in configuration space, at the position of the heavy quark and we will neglect
$O(1/M)$ corrections. The fact that the heavy quarks meet at the origin, i.e.
that their impact parameter is zero means that our model will describe the S
wave channel.

Concerning the Hamiltonian for the antiquarks $\bet$ and $\bal$ we will use a
color potential introduced in \cite{chromo}:

\[
H(t)= \frac {p_\bet^2}{2 m}+\frac {p_\bal^2}{2 m}+
\sum_a\lambda^a_B\lambda^a_\bal V(\vec r_\bal-\vec vt)+
\lambda^a_C\lambda^a_\bal V(\vec r_\bal+\vec vt)+\lambda^a_B\lambda^a_\bet
V(\vec r_\bet-\vec vt)\]
\be
\lambda^a_C\lambda^a_\bet V(\vec r_\bet+\vec vt)+\lambda^a_\bal\lambda^a_\bet
V(\vec r_\bet-\vec r_\bal) + \lambda^a_C\lambda^a_B V(2 \vec vt)
\label{ham}\ee
where $\lambda_C^a$ is the Gell-Mann $SU(3)$ Hermitean matrix applying to quark
$C$, etc, and where $-V(\vec r)$ is a rotation-invariant confining
potential, so that color singlet mesons are
bound together (the $\lambda^a\lambda^a$ factor is negative on a singlet). The
Hamiltonian is bounded from below when restricted  to overall color-singlet
states.

$H(t)$ in (\ref{ham}) is invariant for the Permutation : $P\equiv\bal
\leftrightarrow \bet$. It results that all eigenstates of $H$ will be
eigenstates of $P$ with eigenvalue $\pm 1$.  The asymptotic states, when $T\to
\pm \infty$, are built from simple products of the mesonic wave functions whose
center of mass are located at $\pm \vec v T$:

\be \sqrt 2 |D^{\pm,n,m},T>\eqtipm |M^{(n)}_{C\bet}(-\vec v T)>\otimes
|M^{(m)}_{B\bal}(\vec v T)> \pm |M^{(n)}_{C\bal}(-\vec v T)>\otimes
|M^{(m)}_{B\bet}(\vec v T)> \label{as1}\ee
where the $M^{(n)}(\vec r)$ states are the mesons states defined from eq.
(\ref{meson}) by a Fourier transform on the center of mass variable.
We will assume the evolution in time to be adiabatic, i.e. we assume that
the state $|D^{\pm,n,m},t>$ evolves in time by remaining an eigenstate of
$H(t)$ for all $t$. We will further assume that during the evolution, the
fundamental states $|D^{\pm,0,0},t>$ never cross other states.
It results that the two-dimensional subspace spanned by $|D^{\pm,0,0},t>$ is
stable under the action of the strong Hamiltonian, i.e. that any state within
this subspace evolves  into a state within this subspace at a later time.
Consequently, the two-by-two  restriction of the S-matrix to this subsapce has
to be unitary.
We will now restrict ourselves to the study of the strong interaction
scattering process  of these two fundamental states. At time $t$, the
eigenstates verify:
\be
H(t)|D^{\pm,0,0},t> = E^\pm(vt) |D^{\pm,0,0},t>\ee
where we made use of the fact that $H$ depends on $t$ only through the product
$vt$. The interaction between the two terms in the r.h.s. of (\ref{as1}) is
$O(1/N_c)$ as already argued, see (\ref{colap}). Hence
\be E^+(vt)-E^-(vt) = O(\frac 1 {N_c})\label{ounc}\ee

Asymptotically it is also obvious that
\be E^+(\pm \infty)=E^-(\pm \infty).\ee
since the two mesons do not overlap implying that the two terms in the r.h.s.
of (\ref{as1}) become orthogonal.

In fact the two energies differ from zero only when the two mesons overlap and
the overlap falls off exponentially when $vt\to \infty$.
In the basis $|D^{\pm,0,0},\pm \infty>$, (\ref{as1}), the $S$ matrix is
diagonal and, being unitary its general form writes:
\be S= e^{2i\delta} \left( \begin{array}{cc} e^{2i\phi}& 0\\
0 & e^{-2i\phi}
\end{array}\right)\ee
where
\[ 2\delta= -\int_{-\infty}^{+\infty}\frac{dz}v \left[\frac{E^+(z)+E^-(z)} 2 -
\frac{E^+(\infty)+E^-(\infty)} 2\right]\]
\be 2\phi = -\int_{-\infty}^{+\infty}\frac{dz}v \frac{E^+(z)-E^-(z)} 2.
\label{deph}\ee

Indeed, the S-matrix is given by
\be S=T\exp\left[-i\int^\infty_{-\infty} dt H_I(t)\right]\ee
which in our case of a state, say  $|+>$, that remains eigenstate of a time
dependent Hamiltonian with energy $E^+(t)$, simplifies to
\be S_{++} = \exp\left[-i \int^\infty_{-\infty} dt
(E^+(t)-E^+(\infty)\right]\ee
where the interaction Hamiltonian has been taken to be the total Hamiltonian
minus the
Energy of two non-interacting mesons.

In the meson-meson basis:
\be |M^{(0)}_{C\bet}(-\vec v T)>\otimes
|M^{(0)}_{B\bal}(\vec v T)>\qquad
|M^{(0)}_{C\bal}(-\vec v T)>\otimes
|M^{(0)}_{B\bet}(\vec v T)>\label{bas}\ee
the $S$ matrix writes:
\be S=  e^{2i\delta}  e^{2i\phi \sigma_1}=  e^{2i\delta}
\left(\begin{array}{cc} \cos 2\phi &i \sin 2\phi\\
i \sin 2\phi & \cos 2\phi \end{array}\right)\label{S}\ee

This matrix is, as expected, unitary and invariant for the permutation $P$
which permutes the lines and the columns of the matrix.
When the angle $\phi$ does not vanish, there is a quark exchange between the
mesons which becomes maximal for $\phi=\pi/4$. Actually, $\phi = O(1/N_c)$ from
eq. (\ref{ounc}). From eq. (\ref{deph}), we see that $\phi\propto 1/v$. The
reason for this is clear: the lower the velocity, the longer the mesons overlap
and can exchange quarks. We will now return in our model to the contradiction
discussed in the preceeding section between duality and factorization. We work
out an illustrative example of these features in the Appendix.

\subsection{Final state interaction in our model.}
\label{sec-fsitoy}

In this section we return to the weak interaction. We consider here the
exclusive decay
channels $P_{A\bal}\to M^{(0)}_{B\bal} M^{(0)}_{C\bet}$ and $P_{A\bal}\to
M^{(0)}_{C\bal} M^{(0)}_{B\bet}$. Thanks to the statement made in the
preceeding section that the subspace spanned by these two final states is
stable for the weak interactions, we can safely forget all other channels in
our study of the FSI.

We now assume that the weak Hamiltonian acts at $t=0$, creating the heavy
quarks $B$ and $C$ at $\vec r=0$. The weak interaction creates a state $|f>$ as
defined in (\ref{f}). When $t=0$ an additional symmetry is present in the
strong Hamiltonian  $H(0)$: invariance under permutation of color labels
$C_C\leftrightarrow C_B$ and consequently  under $C_\bet\leftrightarrow
C_\bal$. It results that the $|D^{\pm ,0,0},0>$ states are even (odd) under the
color permutation $P_c: C_C\leftrightarrow C_B$. Restricted to
the color-even (color-odd) sector the Hamiltonian $H(0)$ reduces to:
\be
H(0)= \frac {p_\bet^2}{2 m}+\frac {p_\bal^2}{2 m} -\frac {2(N_c\pm 2)(N_c\mp
1)} {N_c} \left( V(\vec r_\bal)+V(\vec r_\bet)\right)
\pm \frac{2(N_c\mp 1)}{N_c}V(\vec r_\bet-\vec r_\bal)\label{hamred}\ee
where the upper (lower) sign corresponds to color-even (color-odd) states.

It is important to notice that to leading order in $N_c$ the Hamiltonian
(\ref{hamred}) is the same for color-even and color-odd states. It results that
in the $N_c\to \infty$ limit, the two color wave funtion multiply the same
spatial wave function for $t=0$. Furthermore, to leading order in $N_c$ the
Hamiltonian (\ref{hamred}) is equal to twice the Hamiltonian for one
heavy-light meson:
\be H^{(1)}= \frac {p^2}{2 m}- \frac {2(N_c-1)(N_c+1)} {N_c}  V(\vec
r)\label{ham1}\ee
where $p$ and $\vec r$ are the light quark momentum and position. This means
that to leading order in $N_c$ the Hamiltonian (\ref{hamred}) corresponds just
to the sum of two non interacting mesons superposed at the origin. This
corresponds to the factorization assumption.

Eq. (\ref{hamred}) exhibits a symmetry for the exchange of spatial variables
$\vec r_\bal
\to \vec r_\bet$. This symmetry is simply a product of the color permutation
symmetry $P_c$, valid at $t=0$, and the global permutation symmetry $P$, valid
for all $t$.
The eigenstates of (\ref{hamred}) are eigenstates of the spatial permutation
$\vec r_\bal
\to \vec r_\bet$, and it is not difficult to guess that the ground states are
symmetric under the latter permutation. This is illustrated in the Appendix.

Hence, restricting ourselves to the subspace spanned by the two fundamental
states, which are symmetric states for the permutation $\vec r_\bal
\to \vec r_\bet$, we have $P_c=P$ and we project the state $|f>$, (\ref{f}),
into the subspace
\be {\cal H}_0=|D^{+ ,0,0},0>\oplus |D^{-,0,0},0>\label{overlap2}\ee

We have:
\be <D^{\pm ,0,0}0|f>=\frac{(C_1 \pm C_2)(1\pm\frac 1{N_c})}
{(2\pm\frac 2{N_c})^{\frac 1 2}} S^{\pm} \label{col}\ee
where $S^{\pm}$ is the spatial overlap. As stated above $S^\pm=S^0+O(1/N_c)$
where $S^0$ is the spatial overlap of $|f>$ with the direct product of
 two non interacting mesons located at the origin, i.e. $\psi^{(1)}(\vec
r_\bal)\psi^{(1)}(\vec r_\bet)$,
$\psi^{(1)}(\vec r)$ being the ground-state eigenfunction of $H^{(1)}$
(\ref{ham1}).

The evolution forward in time of the states $|D^{\pm ,0,0}, t>$ is obtained by
replacing in (\ref{deph}) the $-\infty$ lower bounds of the integrals by 0. It
results, thanks to time reversal, in phase shifts which are simply divided by
2: $e^{i\delta \pm i \phi}$.

The resulting $T$ matrix for the decay of the initial meson $P_{A\bal}$ into
the fundamental mesons is:
\bea T(P_{A\bal}\to M^{(0)}_{C\bet}M^{(0)}_{B\bal})=&
e^{i\delta}\left\{\frac{(C_1 +C_2)(1+\frac 1 {N_c}) S^+}
{2(1+\frac 1 {N_c})^{\frac 1 2}}\,e^{i\phi}+ \frac{(C_1 -C_2)(1-\frac 1 {N_c})
S^-}
{2(1-\frac 1 {N_c})^{\frac 1 2}}\,e^{-i\phi}\right\}\nn \\
T(P_{A\bal}\to M^{(0)}_{C\bal}M^{(0)}_{B\bet})=& e^{i\delta}\left\{\frac{(C_1
+C_2)(1+\frac 1 {N_c}) S^+}
{2(1+\frac 1 {N_c})^{\frac 1 2}}\,e^{i\phi}- \frac{(C_1 -C_2)(1-\frac 1 {N_c})
S^-}
{2(1-\frac 1 {N_c})^{\frac 1 2}}\,e^{-i\phi}\right\}\label{toy}\eea

To perform a systematic $1/N_c$ expansion, let us first define $\Delta S^\pm$
by
\be S^\pm \equiv S^0 + \frac {\Delta S^\pm}{N_c}\label{deltas}\ee

Then, from (\ref{toy}) we obtain to first order in $1/N_c$:
\bea T(P_{A\bal}\to M^{(0)}_{C\bet}M^{(0)}_{B\bal})=&
e^{i\delta}S^0\big\{\left(C_1 +\frac{C_2}{2N_c}+\frac {C_1(\Delta S^++\Delta
S^-)}{2 N_c S^0}\right)\cos\phi
\nn\\ &+
i\left(C_2 +\frac{C_1}{2N_c}+\frac {C_2(\Delta S^++\Delta S^-)}{2 N_c
S^0}\right)\sin\phi \big\}\nn \\
T(P_{A\bal}\to M^{(0)}_{C\bal}M^{(0)}_{B\bet})=& e^{i\delta}S^0\big\{\left(C_2
+\frac{C_1}{2N_c}+\frac {C_2(\Delta S^++\Delta S^-)}{2 N_c S^0}\right)\cos\phi
\nn\\ &+
i\left(C_1 +\frac{C_2}{2N_c}+\frac {C_1(\Delta S^++\Delta S^-)}{2 N_c
S^0}\right)\sin\phi
\big\}\label{toync}\eea

Comparing (\ref{toync}) with eqs (\ref{over2}) and (\ref{factorization}) we see
that if we take in (\ref{toync}) $\Delta S^\pm=\phi=0$ we recover the
factorization corrected \`a la Shifman, i.e. with eq. (\ref{a1a2s}). The global
phase $\delta$ is not relevant here. Consequently we learn that this factor
$1/2$ \`a la Shifman is completed by two other effects at the same order in
$1/N_c$: i) the $\Delta S^\pm$ which reflect the difference, at the time of the
weak decay, between the total $q q \bar q\bar q$ spatial wave function and the
simple product of the two asymptotic meson wave functions; ii) the phase $\phi$
which reflects strictly speaking the final state interaction. The latter phase
$\phi$, although $1/N_c$ suppressed may become very large when $v\to 0$. For
$\phi=\pi/4$ the factorization becomes grossly wrong since the role of the two
final states is interchanged: The operator multiplying $C_1$ in $H_W$ produces
dominantly the $M^{(0)}_{C\bal}M^{(0)}_{B\bet}$ instead of
$M^{(0)}_{C\bet}M^{(0)}_{B\bal}$ as suggested by factorization, and vice versa
for $C_2$. In other words, the class II decays become dominant over the class I
when
$\phi=\pi/4$.

Furthermore, the dynamical origin of the phase $\phi$ and of the factors
$\Delta S^\pm$ is obviously strongly dependent of the precise nature of the
decay channels considered. This indicates that the effects of $\phi$ and
$\Delta S^\pm$ cannot be incorporated in a phenomenological factorization \`a
la BSW, which assumes a given pair of constants $a_1$ and $a_2$ for all the
decay channels of a meson.

In the appendix we have performed an explicit calculation of $\Delta S^\pm$ and
an estimation of $\phi$ for an harmonic oscillator potential. As can be seen
from (\ref{resds}) and \ref{phires}), $\phi$ is always the dominant $1/N_c$
correction when $\bal$ and $\bet$ are light quarks, and when they are heavy,
$\phi$ still dominates as long as $v m R\le 1$.

Let us now consider a more general case by first introducing spin.

\subsection{Final state interactions of the fundamental pseudoscalar and vector
mesons.}
\label{sec-fsispin}

{}From heavy quark symmetry (HQS) we know that vector and pseudoscalar mesons
are degenerate. HQS also tells us that the heavy quark spin is conserved. This
is of course a trivial consequence of (\ref{ham}), but it is quite general. For
example, in  (\ref{ham}) we might add a term coupling the light quark spins,
such as $\vec\sigma_\bet \cdot \vec\sigma_\bal$, but all terms including the
spins $s_B$ and $s_C$ are $1/M$ suppressed.

Restricting ourselves to a 0 total quark spin, the ground states combine into
four possible asymptotic states:
 $P_{C\bet}P_{B\bal}$, $V_{C\bet}V_{B\bal}$, $P_{B\bet}P_{C\bal}$ and
$V_{B\bet}V_{C\bal}$ (where $P$ stands for pseudoscalar and $V$ for vector).

It is then convenient to use states with a given symmetry for $P_S\equiv
s_B\leftrightarrow s_C$. The relevant combinations are:
\bea P_S \left\{\frac{-|P_{C\bet}P_{B\bal}> + |V_{C\bet}V_{B\bal}>}2\right\}
=&  -\left\{\frac{-|P_{C\bet}P_{B\bal}> + |V_{C\bet}V_{B\bal}>}2\right\}\nn\\
P_S \left\{\frac{-3|P_{C\bet}P_{B\bal}> - |V_{C\bet}V_{B\bal}>}2\right\}
=&  \left\{\frac{-3|P_{C\bet}P_{B\bal}> - |V_{C\bet}V_{B\bal}>}2\right\}
\label{ps}\eea
where $VV'$ stands for $V^0V'^0 -V^+V'^--V^-V'^+$ with $0,+,-$ labelling the
polarization of the vector mesons.

In fact, the first combination in (\ref{ps}) corresponds to $S_{BC}=0$ (total
spin of $B$ and $C$) and the second to $S_{BC}=1$. Using  for large $|T|$ the
notation:
\bea |^1D^{(0,0)}_{C\bet;B\bal},T>=&\left\{\frac{-|P_{C\bet}(-\vec
vT)P_{B\bal}(\vec vT)> + |V_{C\bet}(-\vec vT)V_{B\bal}(\vec vT)>}2\right\}\nn\\
|^3D^{(0,0)}_{C\bet;B\bal},T>=&\left\{\frac{-3|P_{C\bet}(-\vec
vT)P_{B\bal}(\vec vT)> - |V_{C\bet}(-\vec vT)V_{B\bal}(\vec
vT)>}2\right\}\label{not}\eea
the strong Hamiltonian is diagonal in the basis where both $P_S$ and
$P\equiv\bal \leftrightarrow \bet$ are diagonal:
\bea \sqrt 2 |^1D^{\pm,0,0},T>\eqtipm &|^1D^{(0,0)}_{C\bet;B\bal},T> \pm
|^1D^{(0,0)}_{C\bal;B\bet},T>\nn\\
\sqrt 2 |^3D^{\pm,0,0},T>\eqtipm &|^3D^{(0,0)}_{C\bet;B\bal},T> \pm
|^3D^{(0,0)}_{C\bal;B\bet},T>
\label{as2}\eea

The four states in (\ref{as2}) evolve diagonally under the strong Hamiltonian
and lead to four phase shifts given by formulae similar to eq. (\ref{deph}).

Next we make the assumption that $|f>$ defined in (\ref{f}) is odd under $P_S$:
\be P_S|f>=-|f> \label{as}\ee

Relation (\ref{as}) is a consequence of Fierz symmetry whenever $H_W$ is build
up of Fierz-invariant currents, as is the case in eq. (\ref{wh}). The fact that
Fierz symmetry translates into a spin antisymmetry as in (\ref{as}) comes from
the fact that Fierz transformation contains an additional minus sign from
fermion field commutation.
It results that only the states $|^1D^{\pm,0,0},0>$ are produced during weak
decay, i.e. $S_{BC}=0$. The arguments from the beginning of section
\ref{sec-fsitoy} to eq. (\ref{col}) may be repeated, except that due to the
spin asymetry, the $|^1D^{+,0,0},0>$
($|^1D^{-,0,0},0>$ is color-odd (color-even):
\be <D^{\pm ,0,0}0|f>=\frac{(C_1 \mp C_2)(1\mp\frac 1{N_c})}
{(2\mp\frac 2{N_c})^{\frac 1 2}} S^{\pm} \label{col2}\ee
leading to
\bea T(P_{A\bal}\to ^1\!D^{(0,0)}_{C\bet;B\bal})=& e^{i\delta}\left\{\frac{(C_1
-C_2)(1-\frac 1 {N_c}) S^+}
{2(1-\frac 1 {N_c})^{\frac 1 2}}\,e^{i\phi}+ \frac{(C_1 +C_2)(1+\frac 1 {N_c})
S^-}
{2(1+\frac 1 {N_c})^{\frac 1 2}}\,e^{-i\phi}\right\}\nn \\
T(P_{A\bal}\to ^1\!D^{(0,0)}_{C\bal;B\bet})=& e^{i\delta}\left\{\frac{(C_1
-C_2)(1-\frac 1 {N_c}) S^+}
{2(1-\frac 1 {N_c})^{\frac 1 2}}\,e^{i\phi}- \frac{(C_1 +C_2)(1+\frac 1 {N_c})
S^-}
{2(1+\frac 1 {N_c})^{\frac 1 2}}\,e^{-i\phi}\right\}\label{toys}\eea
and to first order in $1/N_c$:
\bea T(P_{A\bal}\to M^{(0)}_{C\bet}M^{(0)}_{B\bal})=&\frac \eta 2
e^{i\delta}S^0\big\{\left(C_1 +\frac{C_2}{2N_c}+\frac {C_1(\Delta S^++\Delta
S^-)}{2 N_c S^0}\right)\cos\phi
\nn\\ &-
i\left(C_2 +\frac{C_1}{2N_c}+\frac {C_2(\Delta S^++\Delta S^-)}{2 N_c
S^0}\right)\sin\phi \big\}\nn \\
T(P_{A\bal}\to M^{(0)}_{C\bal}M^{(0)}_{B\bet})=&\frac\eta 2
e^{i\delta}S^0\big\{-\left(C_2 +\frac{C_1}{2N_c}+\frac {C_2(\Delta S^++\Delta
S^-)}{2 N_c S^0}\right)\cos\phi
\nn\\ &+
i\left(C_1 +\frac{C_2}{2N_c}+\frac {C_1(\Delta S^++\Delta S^-)}{2 N_c
S^0}\right)\sin\phi
\big\}\label{toync2}\eea
where $\eta=+1$ for longitudinal vector mesons and $\eta=-1$ for transverse
vector and pseudoscalar mesons. The difference between the r.h.s of
(\ref{toys}) and the r.h.s. of ({\ref{toy}) comes from the interchange of color
symmetric with color antisymmetric
combinations as apparent when comparing (\ref{col}) and (\ref{col2}).

It is to be noted that the relation between $PP$ and $VV$ production amplitude
is exactly given by the fact that only the $^1D$ combination are created.
Indeed, this relation is a consequence of HQS and would only be corrected if we
considered the $O(1/M)$ corrections.
This relation is not a surprise when one realizes that the conditions
(\ref{m1}) and (\ref{m2}) imply an S-wave dominated decay.

\subsection{Semi-inclusive decay.}

We have found important channel dependent corrections to factorization.
We may still wonder if these corrections ares not washed out when we sum up on
one side all the decay channels $M^{(n)}_{C\bet} M^{(m)}_{B\bal}$ for all $m,
n$, and on the other side all the channels $M^{(n)}_{B\bet} M^{(m)}_{C\bal}$.
This is the aim of this section.

Let us call ${\cal H}^+(t)$ (${\cal H}^-(t)$) the Hilbert space spanned by the
set of states $|D^{+,n,m},t>, \forall n,m $ ($|D^{-,n,m},t>, \forall n,m $)
defined in  eq. (\ref{as1}). ${\cal H}^+(t)$ (${\cal H}^-(t)$) contains the
even (odd) states under the permutation $P\equiv \bal\leftrightarrow \bet$. The
latter commuting with the Hamiltonian $H(t)$, the evolution does not mix the
spaces  ${\cal H}^+(t)$  and ${\cal H}^-(t)$. We shall call $U^+(t_1, t_2)$
($U^-(t_1, t_2)$) the evolution operator in ${\cal H}^+(t)$ (${\cal H}^-(t)$).
$U^\pm(t_1, t_2)$ are unitary.

As already stated in section \ref{sec-fsitoy}, $H(0)$ is also invariant under
the color permutation $P_c: C_C\leftrightarrow C_B$. For $t=0$ the Permutation
$P=P_c P_r$ where
$P_r\equiv p_\bal, s_\bal \leftrightarrow p_\bet, s_\bet$ (remember $\vec r_B
=\vec r_C=0$ for $t=0$). We then decompose $|f>$ both into eigenstates of $P$:

\bea
|f^+>=&\frac {(C_1+C_2)(\delta_{c_C,c_\bet}\delta_{c_B,c_\bal}
+ \delta_{c_C,c_\bal}\delta_{c_B,c_\bet})}{2 N_c }
|f^S_+> +\frac {(C_1-C_2)(\delta_{c_C,c_\bet}\delta_{c_B,c_\bal}
- \delta_{c_C,c_\bal}\delta_{c_B,c_\bet})}{2 N_c }|f^S_->
\nn\\
|f^->=&\frac {(C_1-C_2)(\delta_{c_C,c_\bet}\delta_{c_B,c_\bal}
- \delta_{c_C,c_\bal}\delta_{c_B,c_\bet})}{2 N_c }
|f^S_+>+\frac {(C_1+C_2)(\delta_{c_C,c_\bet}\delta_{c_B,c_\bal}
+ \delta_{c_C,c_\bal}\delta_{c_B,c_\bet})}{2 N_c }|f^S_->\nn\\
\label{fexp}\eea
where $|f^\pm>$ are eigenstates of $P$ with eigenvalue $\pm$, $|f_S>$
contains the spin-space part of the wave function
  (\ref{f}):
\bea |f^S>=&\sum_{\begin{array}{l}\tiny{s_C, s_\bet, c_C, c_\bet,}\\ \tiny
{s_B,s_\bal,c_B,c_\bal}\end{array}}
\int d\vec p_C d\vec p_\bet d\vec p_B d\vec p_\bal \Psi(\vec p_C, s_C, \vec
p_\bet, s_\bet,\vec p_B, s_B, \vec p_\bal, s_\bal)\nn\\  &
|C,\vec p_C, s_C, c_C ; \bet, \vec p_\bet, s_\bet,
c_\bet;B,\vec p_B, s_B, c_B ; \bal, \vec p_\bal, s_\bal,
c_\bal > \label{fs}\eea
which is expanded into the eigenvectors of $P_r$, $|f^S_\pm>$ corresponding to
eigenvalues $P_r |f^S_\pm>=\pm |f^S_\pm>$.

The norm of $|f^\pm>$ is:
\be <f^\pm |f^\pm>= \frac 1 2\left[(C_1\pm C_2)^2 \left(1\pm\frac 1
{N_c}\right)<f^S_+|f^S_+>+(C_1\mp C_2)^2 \left(1\mp\frac 1
{N_c}\right)<f^S_-|f^S_->
\right]\label{norm}\ee
If we define
$<f^S_\pm|f^S_\pm>\equiv K_\pm$, then $K_++K_-=K$ as defined in eq. (\ref{K}).

As stated above:
\be |f^\pm> \in {\cal H}^\pm \label{in}\ee

The evolution toward a large positive time $T$ leads to:

\be |f^\pm(T)>= U^\pm(T,0)|f^\pm> \label{evol}\ee

By unitarity the norm of $|f^\pm(T)>$ equals that of  $|f^\pm>$.

We get
\bea \sqrt 2 T(P_{A\bal}\to D^{\pm,m,n})= & {(C_1\pm C_2)(1\pm\frac 1
{N_c})^{\frac 1 2}}  K_+^{\frac 1 2} \,S_+^{\pm,m,n}
e^{i\phi_+^{\pm,m,n}}+\nn\\ &
 {(C_1\mp C_2)(1\mp\frac 1 {N_c})^{\frac 1 2}}  K_-^{\frac 1 2} \,S_-^{\pm,m,n}
e^{i\phi_-^{\pm,m,n}} \eea
where $S_\pm^{\pm,m,n}$ and $\phi_\pm^{\pm,m,n}$ are real numbers defined by:
\be S_\pm^{\pm,m,n} e^{i\phi_\pm^{\pm,m,n}} = \frac 1
{(<f^\pm(T)|f^\pm(T)>)^{\frac 1 2}} <D^{\pm,m,n},T |f_\pm^\pm(T)> \ee
with
\be f^\pm_\pm(T)=  U^\pm(T,0)|f_\pm^\pm(0)>\ee
$f^\pm_+$ ($f^\pm_-$) being the first (second) terms in the right hand sides of
(\ref{fexp}).

{}From unitarity:
\be \sum_{m,n} \left (S_+^{\pm,m,n}\right)^2+ \left
(S_-^{\pm,m,n}\right)^2=1\ee
which leads to
\be \sum_{\pm,m,n}\left |T(P_{A\bal}\to D^{\pm,n,m})\right|^2=
\left(C_1^2+C^2_2 + 2 \frac {C_1 C_2}{N_c}
\right) \ee
 as expected from (\ref{ff}) and unitarity.

Finally
\bea T(P_{A\bal}\to M^{(m)}_{C\bet}M^{(n)}_{B\bal})=& K_+^{\frac 1 2}
\left\{\frac{(C_1 +C_2)(1+\frac 1 {N_c}) S_+^{+,m,n}}
{2(1+\frac 1 {N_c})^{\frac 1 2}}\,e^{i\phi_+^{+,m,n}}+ \frac{(C_1 -C_2)(1-\frac
1 {N_c}) S_+^{-,m,n}}
{2(1-\frac 1 {N_c})^{\frac 1 2}}\,e^{i\phi_+^{-,m,n}}\right\}\nn \\& K_-^{\frac
1 2} \left\{\frac{(C_1 -C_2)(1-\frac 1 {N_c}) S_-^{+,m,n}}
{2(1-\frac 1 {N_c})^{\frac 1 2}}\,e^{i\phi_-^{+,m,n}}+ \frac{(C_1 +C_2)(1+\frac
1 {N_c}) S_-^{-,m,n}}
{2(1+\frac 1 {N_c})^{\frac 1 2}}\,e^{i\phi_-^{-,m,n}}\right\}\nn \\
T(P_{A\bal}\to M^{(m)}_{C\bal}M^{(n)}_{B\bet})=& K_+^{\frac 1 2}
\left\{\frac{(C_1 +C_2)(1+\frac 1 {N_c}) S_+^{+,m,n}}
{2(1+\frac 1 {N_c})^{\frac 1 2}}\,e^{i\phi_+^{+,m,n}}- \frac{(C_1 -C_2)(1-\frac
1 {N_c}) S_+^{-,m,n}}
{2(1-\frac 1 {N_c})^{\frac 1 2}}\,e^{i\phi_+^{-,m,n}}\right\}\nn \\& K_-^{\frac
1 2}
\left\{\frac{(C_1 -C_2)(1-\frac 1 {N_c}) S_-^{+,m,n}}
{2(1-\frac 1 {N_c})^{\frac 1 2}}\,e^{i\phi_-^{+,m,n}}- \frac{(C_1 +C_2)(1+\frac
1 {N_c}) S_-^{-,m,n}}
{2(1+\frac 1 {N_c})^{\frac 1 2}}\,e^{i\phi_-^{-,m,n}}\right\}\nn \\
\label{toyt}\eea

Duality is fully verified since summing over all states
\be \sum_{m,n} \left |T(P_{A\bal}\to M^{(m)}_{C\bet}M^{(n)}_{B\bal})\right|^2
+ \left |T(P_{A\bal}\to M^{(m)}_{C\bal}M^{(n)}_{B\bet})\right|^2 =
\left(C_1^2+C^2_2 + 2 \frac {C_1 C_2}{N_c}
\right) K. \ee

let us now consider the partially inclusive sums:

\bea \sum_{m,n} \left |T(P_{A\bal}\to
M^{(m)}_{C\bet}M^{(n)}_{B\bal})\right|^2\equiv
& \Sigma_{C\bet;B\bal}  \nn\\
\sum_{m,n} \left |T(P_{A\bal}\to M^{(m)}_{C\bal}M^{(n)}_{B\bet})\right|^2\equiv
&
\Sigma_{C\bal;B\bet}
\eea

Nothing general can be said. In the $N_c\to \infty$ limit,
\be S_\pm^{+,m,n}\simeq S_\pm^{-,m,n};\qquad \phi_\pm^{+,m,n}\simeq
\phi_\pm^{-,m,n}\label{sphi}\ee

Assuming that for {\it finite} $N_c$ we keep the relations (\ref{sphi}), we
recover Shifman's Ansatz:
\bea  \Sigma_{C\bet;B\bal}=& \frac K 4\left\{(C_1-C_2)(1-\frac 1 {N_c})^{\frac
1 2}
+(C_1+C_2)(1+\frac 1 {N_c})^{\frac 1 2}\right\}^2 =  K \left\{C_1^2 +
\frac{C_1C_2}{N_c} + O(\frac 1 {N_c^2})\right\}\nn\\
\Sigma_{C\bal;B\bet}=& \frac K 4\left\{(C_1-C_2)(1-\frac 1 {N_c})^{\frac 1 2}
-(C_1+C_2)(1+\frac 1 {N_c})^{\frac 1 2}\right\}^2 = K \left\{C_2^2 +
\frac{C_1C_2}{N_c} + O(\frac 1 {N_c^2})\right\}\nn\\
\eea

If we made the opposite assumption that the relative phases
$\phi_\pm^{+,m,n}-\phi_\pm^{-,m,n}$ are random, which might be reasonable at
small velocity, the result would be
\be \Sigma_{C\bet;B\bal}= \Sigma_{C\bal;B\bet} = K \left\{\frac {C_1^2+C_2^2}2
+ \frac {C_1C_2}{N_c}\right\}\ee
that is, an equal sharing of the total probability between the two sets of
channels.
In the latter case, even though we consider an inclusive sum, the final state
interaction has a non trivial effect: the quarks have been redistributed at
random between the final mesons.
Of course such a random phase equal sharing may only happen when phase space
allows
for many final states to add up in a random way.

As a side remark we would like to mention another, not yet published, study
that we have performed on duality versus factorization.
We have considered a model with non-relativistic scalar quarks bound to color
singlets by a color
harmonic oscillator potential \cite{chromo} without assuming heavy quarks
neither using an adiabatic approximation as done in the present paper. This
model also
automatically restores duality i.e. the conservation of probability. Summing
over all mesons in
the limit in which the radius $R \to \infty$ one should find the free quark
result. This is indeed the
case:
\bea
&& \Sigma_{C\bet;B\bal} \propto  \left[ C_1 + y
{C_2 \over  N_c}\right]^2 + O\left( \frac 1 {N_c^2}\right) \\
&&\Sigma_{C\bal;B\bet} \propto  \left [ C_2 + (1
- y) {C_1 \over N_c} \right ]^2	+ O\left( \frac 1 {N_c^2}\right)\label{jc}
\eea

\noi with $y$ depending non-trivially on the
masses, unlike the universal factor 1/2 proposed by Shifman. It is seen that
FSI restores automatically duality. Notice by the way that
expression (\ref{jc}), for an arbitrary value of $y$, is more general than
Shifman's Ansatz and would as well restore duality.

\section{Conclusions.}

We have used a Quark Model where the motion of heavy quarks is treated as
classical and where we assume two mass-degenerate antiquarks. We have used the
resulting permutation symmetry to simplify the problem. We have restricted
ourselves to the $q q \bar q \bar q$ sector,
and we have shown that the contradiction between standard factorization and
duality
stems from the non-orthogonality in color space of the two decomposition of the
$q q \bar q \bar q$ singlet into two pairs of $q \bar q$ color singlets. Taking
care to use an orthogonal basis that diagonalizes the Hamiltonian,   the
dynamics of this sector shows very clearly how the final state interaction
corrects standard factorization such as to satisfy duality.

Shifman \cite{shifman} has proposed to correct factorization by a replacement
of $1/N_c$ by $1/(2N_c)$ while keeping the phenomenological factorization \`a
la BSW. We have shown that this effect is indeed present. However we find two
additional effects to same order in $1/N_c$. One is related to the spatial
distorsion of the meson wave functions at the time of the weak decay: the two
mesons overlap in space and hence interact strongly. This has been expressed by
our parameter $\Delta S^\pm$.
The second and more important additional effect is the phase difference $\phi$
between the permutation-even and permutation-odd states when evolving after the
decay until the mesons are spatially distant. The latter effect is $O(1/N_c)$
but also $O(1/v)$ where $v$ is the final meson velocity in the total rest
frame, and also $O(1/mR)$ where $m$ is the light antiquarks constituant mass
and $R$ is the wave function radius.  This phase shift effect should dominate
in the small velocity regime for light antiquarks. For large velocity and for
heavy antiquarks, it vanishes.

We have seen that in the exclusive case, restricting ourselves to the ground
state mesons, the phase shift $\phi$ is the dominant $1/N_c$ correction for
small velocity and light antiquarks, and it
 may produce a total modification of the factorization assumption, which could,
for  $\phi=\pi/4$, be large enough to
totally interchange the amplitudes of the two channels, and lead to a dominant
class II decay. In an illustrative example treated in the appendix the $t=0$
wave function distortion , $\Delta S^\pm$ turns out to be small. We ignore if
this is a general feature. Would it be so, it would indicate a validity of
Shifman's Ansatz for large velocity and rather heavy antiquarks.

In the semi-inclusive case we compare the total decay probability into two sets
of channels that correspond to the two possible pairing of $q q \bar q \bar q$
into two $q \bar q$.
Again, the small phase shift, small distortion limit amounts to Shifman's
Ansatz, while the opposite, random phase shift limit,
amounts to an equal sharing of inclusive decay probability into the two sets of
final channels. Again the latter situation may be reasonable in the small
velocity case provided many final states are kinematically allowed.

\section*{Acknowledgements.}

This work was supported in part by the CEC Science Project SC1-CT91-0729 and by
the Human Capital
and Mobility Programme, contract CHRX-CT93-0132.

\appendix
\section{An illustrative example.}

As an illustration of the section \ref{sec-fsitoy},
let us take for the potential $V$ in (\ref{ham}) an harmonic oscilator
potential:
\be V(\vec r)= - \frac{ N_c}
{4 (N_c^2-1) m R^4}\vec r \,^2\label{ho}\ee
so that the ground state solution of (\ref{ham1}) is:
\be \psi^{(1)}(\vec r)=\frac 1 {R^{\frac 3 2}\pi^{\frac 3 4}}e^{- \frac
{r^2}{2R^2}}\label{psi1}\ee
where $r=|\vec r|$.

For the spatial part of the wave function $|f>$ we take in configuration space:
\be \Psi(\vec r_C, \vec r_\bet, \vec r_B, \vec r_\bal)= G \psi^{(1)}(\vec
r_\bal)
\delta_3(\vec r_C) \delta_3(\vec r_B) \delta_3(\vec r_\bet)\label{psif}\ee
which expresses the fact that the weak operator is local\footnote{Remember that
we assume here spinless quarks.} and that the quark $\bal$ is a spectator
coming from the $P_{A\bal}$ meson, i.e. in the ground state wave function. $G$
is proportional to the Fermi constant.

To compute the spatial overlaps $S^\pm$ in (\ref{col}), we need to know the
ground state solutions of the Hamiltonian (\ref{hamred}) with (\ref{ho}) for
$V$. Let us change variables in (\ref{hamred}):
\bea \vec r_\bet= \vec R - \frac 1 2 \vec r,&\qquad \vec r_\bal= \vec R + \frac
1 2 \vec r,\nn \\
\vec p_\bet= \frac 1 2 \vec P -\vec p ,&\qquad p_\bal= \frac 1 2 \vec P +\vec
p,
\eea
leading to
\be H(0)= \frac {P^2}m +\frac {p^2}{4m} +\frac {(N_c\pm2)(N_c\mp 1)}
{4(N_c^2-1)m R^4} \vec R\,^2 +\frac {(N_c \pm 4)(N_c\mp 1)}
{(N_c^2-1)m R^4} \vec r\,^2.\label{Hho}\ee

The ground state solution, i.e. the spatial wave functions of $|D^{\pm,0,0},0>$
is:
\be
\psi^{\pm,0,0}(r_\bal, r_\bet)=\frac {(N_c\pm2)^{\frac 3 8}(N_c\pm4)^{\frac 3
8}}
{(N_c\pm 1)^{\frac 3 4}R^{3}\pi^{\frac 3 2}}e^{-\frac{(N_c\pm 2)^{\frac 1 2}
(\vec r_\bal+\vec r_\bet)^2}{4(N_c\pm 1)^{\frac 1 2}R^2}}
e^{-\frac{(N_c\pm 4)^{\frac 1 2}
(\vec r_\bal-\vec r_\bet)^2}{4(N_c\pm 1)^{\frac 1 2}R^2}}\ee

The overlap is given by
\bea S^\pm=&
\int d^3r_\bal d^3r_\bet G \psi^{(1)\dagger}(\vec r_\bal)
 \delta_3(\vec r_\bet)\psi^{\pm,0,0}(r_\bal, r_\bet)\nn \\
=&
G \frac { 4^{\frac 3 2}(N_c\pm2)^{\frac 3 8}(N_c\pm4)^{\frac 3 8}}
{\left[(N_c\pm2)^{\frac 1 2}+(N_c\pm4)^{\frac 1 2}+2 (N_c\pm 1)^{\frac 1 2}
\right]^{\frac 3 2} R^{\frac 3 2}\pi^{\frac 3 4}}\label{spm}\eea
where we have in (\ref{psif}) left aside the $\delta$ functions related to the
heavy quarks, since the latter are treated  classically. One has also for the
factorization hypothesis result:
\be S^0= G\frac 1 {R^{\frac 3 2}\pi^{\frac 3 4}}\label{s0}\ee
which is the $\psi^{(1)}(0)$ coming from the overlap of $\delta(\vec r_\bet)$
with
$\psi^1(\vec r_\bet)$. The overlap of $\psi^{(1)}(\vec r_\bal)$ with itself
gives obviously 1. $S^0$ in (\ref{s0}) is obviously the $N_c\to\infty$ limite
of $S^\pm$ in  (\ref{spm}).

To next to leading order in $1/N_c$ the calculation of $\Delta S^\pm$ defined
in (\ref{deltas}) is now straightforward from (\ref{spm}) and (\ref{s0}):
\be \frac {\Delta S^\pm}{N_c} = \pm \frac 3 {8 N_c}\label{resds}\ee
which turns out to be rather small, mainly because the normalization factor
compensates for a large part the modification of the integral.

{}From (\ref{Hho}) we also get the ground state energy:
\be E^\pm(0)= \frac 3 {2 m R^2} \left\{\left(\frac {N_c\pm 2}{N_c\pm
1}\right)^{\frac 1 2}+\left(\frac {N_c\pm 4}{N_c\pm 1}\right)^{\frac 1 2}
\right\}\label{epmho}\ee
leading to
\be E^+(0)-E^-(0)= \frac 6 { m R^2 N_c} + O\left(\frac 1
{N_c^2}\right)\label{diffe}\ee

To compute $\phi$ we need, (\ref{deph}), to know $E^+(z)-E^-(z)$ for all $z\ne
0$. This is not so easy to compute. We will simply use (\ref{diffe}) to make an
order of magnitude estimate. We will assume $E^+(z)-E^-(z)$ to be equal by
$E^+(0)-E^-(0)$ as long as the hadrons overlap, i.e. for $|z|\le c R$ where
$c$ is some number of order 1\footnote{The wave function radius for the wave
function (\ref{psi1}) is $<\vec r\,^2>=3/2 R^2$. Hence one might think of
taking $c\simeq 3/2$.}. Then
\be
\phi \sim \frac {3 c}{m v R N_c}. \label{phires}\ee

Although purely indicative, this result, besides confirming that $\phi \propto
1/v  N_c$ also learns us that $m R$ is the dimensionless number that gives the
scale. When $\bal$ and $\bet$ are light quark, it is known that $mR\sim 1$.
Comparing (\ref{phires}) with (\ref{resds}) and to the $1/(2N_c)$ correction
\`a la Shifman, we check that in the exclusive case, the phase shift $\phi$ is
the dominant $1/N_c$ contribution.

\end{document}